\documentstyle[aps,epsf,epsfig,prl,multicol]{revtex}
\begin{document}
\draft

\title{ On the Nonlinear Kr\"onig-Penney model }
\author{WeiDong Li$^{1,2}$, A. Smerzi$^{1,3}$}
\address{
$^1$Istituto Nazionale per la Fisica della Materia BEC-CRS
and Dipartimento di Fisica, Universit\`{a} di Trento, 
I-38050 Povo, Italy\\
$^2$Department of Physics and Institute of Theoretical Physics, 
Shanxi University, Taiyuan 030006, China\\
$^3$Theoretical Division, Los Alamos National Laboratory, 
Los Alamos, NM 87545, USA\\ 
}

\maketitle

\begin{abstract}
We study the nonlinear Schr\"odinger equation with a periodic
delta-function potential. This realizes a nonlinear 
Kr\"onig-Penney model, with physical applications in the context of
trapped Bose-Einstein condensate alkaly gases and in the transmission of 
signals in optical fibers. 
We find analytical solutions of zero-current Bloch states. Such
wave-functions have the same periodicity of the 
potential, and, in the linear limit, reduce to the 
Bloch functions of the Kr\"onig-Penney model.
We also find new classes of solutions having a periodicity different
from that of the external potential. 
We calculate the chemical potential of
such states and compare it with the linear excitation spectrum.
\end{abstract}
\date{today}
\begin{multicols}{2}

Nonlinearity can deeply 
modify the Bloch theory of non-interacting atoms trapped in 
periodic potentials.
Loop structures,
energetic and dynamical instabilities,
solitons and ``generalized" Bloch states (i.e. states which do not
share the same periodicity of the lattice), all arise in the context of 
a nonlinear Schr\"odinger (or Gross-Pitaevskii) equation. Applications
span, for instance, the physics of
dilute Bose-Einstein condensed gas trapped in 
optical lattices 
\cite{bronski01,niu1,niu2,muller,moler,Pethick,pethick1,stringari,stringari1}
or the propagation
of signals in optical fibers \cite{optical}

In this paper we study analitically 
the nonlinear Schr\"odinger
equation with an external Kr\"onig-Penney (KP) potential, 
given by a periodic array of delta-functions.
The linear Schr\"odinger equation with the same potential
has been solved quite early in the '30,
playing a distinguished role as a model in metal's theory
\cite{hennig99,kittler}.

It is noticeable that also several properties of the
nonlinear Schr\"odinger equation, with the
same KP external potential, can be derived analitically. The most interesting  
results, however, are related with the emergence of new properties
which do not have a counterpart in the linear case.
Stationary solutions 
of the GPE which do not reduce to any of the
eigenfunctions for a vanishing nonlinearity have been studied 
in \cite{Pethick} using a tight binding approximation, in and two-wells
systems, in \cite{Reinhardt,Reinhardt1,Reinhardt2,presilla,raghavan99}. 

The mean-field model of a quasi-1D BEC trapped in a KP potential
is governed by the following nonlinear Schr\"odinger (or Gross-Pitaevski) 
equation: 
\begin{equation}
\mu \psi \left( x\right) =\left( -\frac{\hbar ^2}{2m}\frac{\partial ^2}{%
\partial x^2}+V\left( x\right) +g \mid \psi \left( x\right) \mid ^2\right)
\psi \left( x\right)  \label{g1}
\end{equation}
where $\mu$ is the chemical potential and $g$ 
the nonlinear coupling constant. The KP external potential
is given by equispaced delta-functions:
$V(x) = p~\sum^{\infty}_{n=-\infty}~\delta(x - n a)$, 
having a lattice constant $a$.
Since the external potential has a step-like shape,
it is useful to rewrite the GPE in hydrodynamic form. With
$\psi \left( x\right) =\sqrt{
\rho \left( x\right) }\exp \left[ -i\Theta \left( x\right) \right]$ 
(and in dimensionless units), we have: 
\begin{eqnarray}
&&\left( {{\partial \rho}\over {\partial x}} \right)^2 = 
2 \eta \rho^3 + 
4 (\mu - V) \rho^2 - \beta \rho - 4 \alpha^2 \cr
&& \Theta = \int dx {\alpha \over \rho}
\label{hydro}
\end{eqnarray}
with $\eta = g N_0 ({2 m a^2} / \hbar^2)$,
${x \over a} \to x$, $(\mu,V)~ {{2 m a^2} / \hbar^2} \to (\mu,V)$, 
$a \rho \to \rho$ and
normalization $\int_n^{n\pm 1} d x~ \rho(x) = 1$. $N_0$ is the number of 
atoms in each well, and the 
integration constants $\alpha, \beta$ are fixed by the 
boundary conditions.
In particolar, $\alpha $ has a simple physical meaning, being 
the current carried by the order parameter $\psi(x)$:
\begin{equation}
J = \alpha
\label{current}
\end{equation}
(where $J {a^2 \over \hbar} \to J$).

{\em Bloch state.} We derive a class of stationary solutions of
Eq.s(\ref{hydro}) having (i) the same periodicity of the external potential
(Bloch states), and (ii) the linear limit: 
such states reduce to the 
well know solutions of the linear KP model when $\eta \to 0$.
The Bloch theorem assures the possibility to write the complete set
of eigenstates of the linear equation in the form:
$\psi \left( x\right) =\exp \left( iqx\right) f_q\left( x\right) $,
where $q$ is the quasi-momentum of the state 
and $f_q\left( x\right)$ is periodic with the lattice constant,
$f_q\left(x + a\right)=f_q\left( x\right)$. 
It is immediate to extend the Bloch theorem and show that stationary states 
having the same periodicity of the potential exist also in
the nonlinear case. However, there are new
classes of states 
which do not have such properties, see next section.

Bloch nonlinear stationary states of the GPE with the KP external 
potential can be written in terms of the Jacobi Elliptical functions. 
With the same class of functions
have been previously studied the exact
stationary solutions of the GPE descibing a train of solitons 
\cite{gurevich}, or the stationary solutions of the GPE in a double 
square-well potential \cite{Reinhardt,Reinhardt1,Reinhardt2}.  
To the best of our knowledge, however, the nonlinear Schr\"odinger
equation with a periodic KP potential
has not been studied so far. We notice that our results with the delta-function
potential can be generalized (although becomes rather 
cumbersome) to a regular array of square-wells, as will been shown elsewhere. 

Stationary solutions of the Eq.s(\ref{hydro}) are:
\begin{equation}
\rho ( x) =\frac A{8K^2}-( \frac A{8K^2}-\frac{128K^4\alpha
^2}{A( 16K^4+A\eta ) }) \text{SN}^2( Kx+\delta
,n) ,  \label{f1}
\end{equation}
where 
\begin{eqnarray}
n &=& -\frac A{16K^4}\eta +
\frac{64K^2\alpha ^2}{A\left( 16K^4+A\eta \right)}\eta \cr
\beta &=& - \frac{ (16 K^4 A + A^2 \eta)^2 + 2048 (A \eta + 8 K^4) 
K^6 \alpha^2}{32 K^4 (16 A K^4 + A^2 \eta) } \label{par} \cr
\mu &=& K^2+\left( \frac A{8K^2}+
\frac{64K^4\alpha ^2}{A\left( 16K^4+A\eta \right) }\right) \eta
\end{eqnarray}
and SN$\left( u,n\right)$ is the Jacobian elliptic sine function. 
SN$\left( u,n\right)$ has the desiderable property to give, 
in the linear limit, $\eta = 0$,
$\rho \left( x\right) = \frac A{8K^2}\cos ^2\left( Kx+\delta \right) +
8\frac{\alpha ^2}A\sin^2\left( Kx+\delta \right)$, 
$\Theta(x) = \arctan(8 K \alpha \tan(K x + \delta)/A )$,
which are the exact solutions of the linear KP model.

The three parameters $A$, $K$ and $\delta $
are fixed by imposing the continuity 
of the order parameter and the Bloch periodicity, as in the linear case.
Because of the nonlinearity, however, we need one more condition to fix the
chemical potential: the normalization of the density.
We therefore obtain four conditions: 
\begin{eqnarray}
&&\rho _1\left( 0\right) =  \rho _1\left( 1\right) \cr
&&\partial _x\rho _1\left( 0\right) -\partial
_x\rho _1\left(1 \right) = 2P\rho _1\left( 0 \right) \cr
&&\Theta _1\left(0\right) -
\left( \Theta _1\left( 1 \right) -qt\right) = 2n\pi \cr
&&\int_0^1\rho_1\left( x\right) dx=1
\label{mc}
\end{eqnarray} 
In this paper we consider the special case of zero-current states, 
having $\alpha =0$ (so that $\Theta = const.$).
Using the elementary properties of the
Jacobian Elliptical function\cite{hand} the first two equations 
of (\ref{mc}) give:
\begin{eqnarray}
&&\text{CN}\left( K,n\right) +
\frac P{2K}\text{SN}\left( K,n\right) =\pm 1 \cr
&&n= {{K^2-\mu}\over {2 K^2}}
\label{il}
\end{eqnarray}
%\begin{equation}
%\text{CN}\left( K, {{K^2 - \mu}\over {2 K^2}}  \right) +
%\frac P{2K}\text{SN}\left( K, {{K^2 - \mu}\over {2 K^2}}  \right) =\pm 1
%\label{il}
%\end{equation}
%with $n=\frac 12\left( 1-\mu /K^2\right)$.
In the linear limit such states are at 
the top or at the bottom of the corresponding energy bands
(which is not necessarly true in the nonlinear case
when loop-like structures appear in the
excitation spectrum).
Indeed, the Eq.(\ref{il}) with $\eta = 0$ (giving $\mu = K^2$)
reduces to the well known relation
\begin{equation}
\cos K +\frac P{2K}\sin K =\pm 1
\label{illinear}
\end{equation}
giving the band and the gap widths in the linear K-P model \cite{kittler}. 

To solve Eq.(\ref{il}) we need a further relation between the chemical 
potential $\mu$ and $K$, provided by the normalization in (\ref{mc}).
We obtain two conditions associated with the $+1$ and the $-1$ 
of Eq.(\ref{il}), respectively: 
\begin{eqnarray}
&&\frac \eta {2K}+2{\cal E}\left( Am\left( \frac{K}2,n\right) ,n\right)
-\left( 1-n\right) K =0~ \text{ (+)} \cr
&&\frac \eta {2K}+{\cal E}\left( Am\left( \varphi _2,n\right) \right) -{\cal E}%
\left( Am\left( \varphi _1,n\right) \right) + \cr
&&~~~~~~~~~~~~~~~~~~~~~~~~~~~~~~~~~~~~~~~~-\left( 1-n\right) K =0~\text{(-)}
\label{norm}
\end{eqnarray}
where $n=\frac 12\left( 1-\mu /K^2\right)$,
$\varphi _2={\cal K}\left( n\right) +\frac{K}2$ and 
$\varphi _1={\cal K}\left( n\right) -\frac{K}2$. 
${\cal E}\left( u,n\right) $ is the
incomplete elliptic integral of the second kind and ${\cal K}$ is the
complete elliptic integral of the second kind. $Am\left( \phi_i,n\right) $ 
is the amplitude of $\phi_i$. 

The coupled equations (\ref{il}) and (\ref{norm}) can be solved graphically
as shown in Fig.(1). The black lines are solutions of (\ref{il}), while
the color lines are solution of (\ref{norm})) 
(the dashed and the continuum lines correspond to the $+,-$ sign, 
respectively).
The lines among the intersections between the black and the
color lines (evidenciated by the color filled regions), give
the values of ($\mu, K$) corresponding to the zero-current Bloch states. 
As expected, nonlinearity quite modifies 
the lower energy bands of the systems,
while, for higher bands, differences are reduced. 
\begin{figure}
\centering
\psfig{file=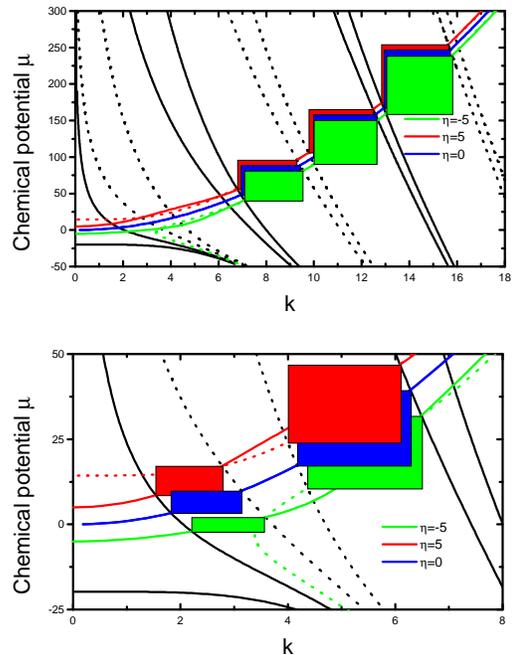, scale=0.5, bbllx=70, bblly=100, bburx=575, bbury=620}
\caption{\label{fig1} Graphical solution of the system of Eq.s(\ref{il}) and
Eq.s(\ref{norm}) for different values of nonlinearity. 
The allowed values of the 
chemical potential $\mu$ as a function of momentum $K$ are given by the 
intersections between the black and the color lines, 
in the color filled regions}
\end{figure}

{\em Generalized Bloch states.}
In the previous section we have considered Bloch-like solutions, where
the condensate wavefunction have the same periodicity 
of the potential. As already mentioned, the nonlinear interaction
allows for stationary solutions which can have, in principle, any
integer value period. In the following, we only consider
solutions which are periodic every two sites: $f_q(x+2) = f_q(x)$.
Notice that wave functions with $q=0,\pi $ 
are symmetric, $\psi \left( x\right) =\psi \left( x+2\right) $, 
while are antisymmetric when 
$q=\pi /2$, $\psi \left(x\right) =-\psi \left( x+2\right) $.
Therefore, $f_{q=0}(x) = f_{q=\pi}(x)$, and the respective chemical potentials
are equal.

In the following, we consider as elementary
cell two neighboring wells separated by the delta potential, with
$\rho _l\left( x\right)$ and  $\rho _r\left(x\right)$ being 
the densities in the ``left" and ``right" well, respectively. 
As in the previous section, we 
consider only states with zero current $\alpha = 0$.
The continuity and the periodicity conditions give:
\begin{eqnarray}
&&\rho _l\left( 0\right) =\rho _r\left( 2 \right) \cr
&&\partial _x\rho _l\left( 0\right) -\partial _x\rho _r\left( 2 \right)
=2P\rho _l\left( 0\right) \cr
&&\rho _l\left( 1 \right) =\rho _r\left( 1 \right) \cr
&&\partial _x\rho _r\left( 1 \right) -\partial _x\rho _l\left( 1 \right)
=2P\rho _l\left( 1 \right) \cr 
&& \int_0^1\rho _l\left( x\right)dx + \int_1^{2}\rho _r\left( x\right) dx= 2\cr
&&K_1^2+\frac{A_1}{8K_1^2}\eta-K_2^2-\frac{A_2}{8K_2^2}\eta =0
\label{mc1}
\end{eqnarray}
First, let's consider the solution which have nodes at the boundary of the elementary cell: 
\begin{equation}
\rho _l\left( 0\right) = \rho_r\left( 2\right) = 0.
\label{3nodes}
\end{equation}
Then we have $\delta _1=\left( 2l_1+1\right) {\cal K}\left( n_1\right) $, $%
\delta _2+2K_2=\left( 2l_2+1\right) {\cal K}\left( n_2\right) $. 
Notice that 
the boundary conditions are equal to those of a double well, 
except for the condition on the first derivativies at the borders of
the elementary cell, which gives 
$A\left( 1-n_1\right) =B\left( 1-n_2\right) $. 
Combining with the normalization,
we arrive at $A=B$ and $K=Q$. Notice that this relation 
excludes the existence of symmetry broken solutions which are
instead obtained 
in a single double-well potential \cite{Reinhardt1}. In particolar, when
$\rho _{l,r}\left( 1\right) =0$, the only solutions are 
the Bloch states. 

We now consider the case  $l_1=l_2=0$, which gives:
\begin{eqnarray}
-2K\text{SC}(K+{\cal K}\left( n\right) )\text{DN}(K+{\cal K}\left( n\right)
)+P &=&0 \\
\frac A{8K^2}\int_0^1\text{CN}^2\left( Kx+{\cal K}\left( n\right) \right) dx
&=&1
\end{eqnarray}
In Fig.2, we present solutions which are symmetric with respect to
the axis $x=1$. These have
a linear limit, which is simply given by the
superposition of two eigenfunctions having the same energy but opposite 
momentum. Obviously, in the nonlinear case
the superposition principle 
breaks down, so that the existence of such solutions was not obvious.
\begin{figure}
\centering
\psfig{file=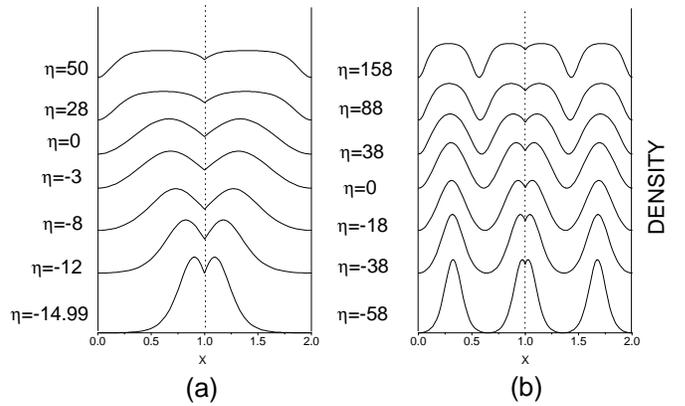, scale=0.5, bbllx=30, bblly=250, bburx=800, bbury=600}
\caption{\label{fig2} Symmetric generalized Bloch states with 
no nodes (a), and with two nodes (b).}
\end{figure}
Solutions with a node in $x=1$ can be constructed from
the previous conditions and 
imposing $\rho _l\left( x\right) =\rho _r\left( x+1\right) $, 
$\rho_r\left( x\right) =\rho _l\left( x-1\right) $. 

There is a different class of generalized Bloch states 
having nodes located outside the boundaries 
of the potential. The conditions are:
\begin{eqnarray}
&&\frac{A_1}{8K_1^2}\text{CN}^2\left( \delta _1\right) =\frac{A_2}{8K_2^2}
\text{CN}^2\left( 2K_2+\delta _2\right) \cr
&&K_2\text{SC}\left( 2K_2+\delta _2\right) \text{DN}\left( 2K_2+\delta
_2\right) + \cr
&&~~~~~~~~~~~~~~ -K_1\text{SC}\left( \delta _1\right) \text{DN}\left( \delta
_1\right) =P \cr
&&\frac{A_1}{8K_1^2}\text{CN}^2\left( K_1+\delta _1\right) =
\frac{A_2}{8K_2^2}
\text{CN}^2\left( K_2+\delta _2\right) \cr
&&K_1\text{SC}\left( K_1+\delta _1\right) \text{DN}\left( K_1+\delta
_1\right) + \cr
&&~~~~~~~~~~~~~~ - K_2\text{SC}\left( K_2+\delta _2\right) \text{DN}\left(
K_2+\delta _2\right) =P \cr
&&K_1^2+\frac{A_1}{8K_1^2}\eta -K_2^2-\frac{A_2}{8K_2^2}\eta =0 \cr
&&\frac{A_1}{8K_1^2}\int_0^1\text{CN}^2\left( K_1x+\delta _1\right) dx+ \cr
&&~~~~~~~~
+\frac{A_2}{8K_2^2}\int_1^{2}\text{CN}^2\left( K_2x+\delta _2\right) dx=2
\end{eqnarray}
The density profiles of such solutions are shown in the Fig.(3), with different
numbers of nodes.
\begin{figure}
\centering
\psfig{file=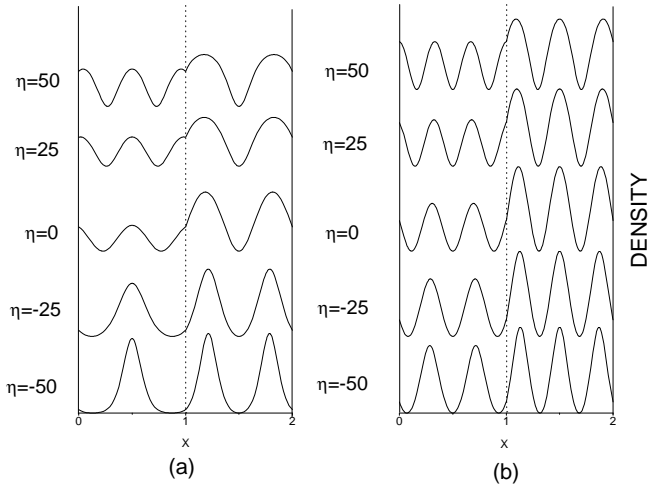, scale=0.5,  bbllx=30, bblly=210, bburx=675, bbury=600}
\caption{\label{fig3} Symmetry broken generalized Bloch states
with three nodes (a) and with five nodes (b).}
\end{figure}
In Fig. 4 we plot the chemical potential of the system as a
function of nonlinearity. 
The full lines corresponds to zero-current
Bloch states, while the dashed and dotted lines correspond to generalized
Bloch states. 
Notice that when $\eta \simeq -8$,
the ground state ($q=0$, Bloch state) is replaced 
by a symmetry broken Bloch state.
\begin{figure}
\centering
\psfig{file=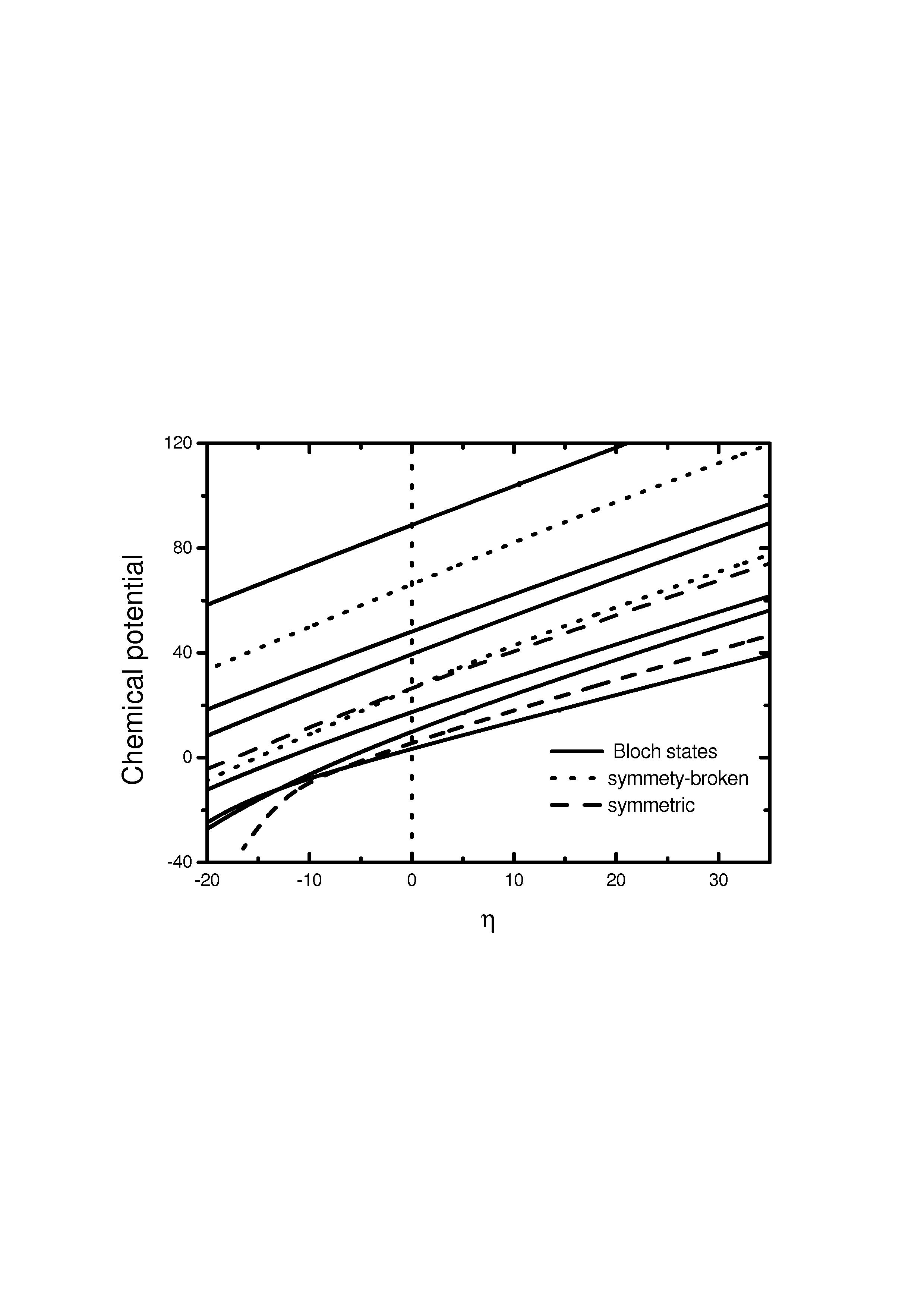, scale=0.5, bbllx=50, bblly=210, bburx=575, bbury=580}
\caption{\label{fig5} Chemical potential as a function of nonlinearity}
\end{figure}

{\em Conclusions}
We have studied the nonlinear Kr\"onig-Penney model. This is given by the
nonlinear Schr\"odinger equation with a periodic
delta-function external potential.
We ve found analytical solutions of zero-current states having 
the same (Bloch states) or a different (``generalized" Bloch states) 
periodicity of the potential. 
Nonlinear Bloch states  reduce, in the linear limit, to the
well known eigenfunctions of the linear Kr\"onig-Penney model.
We have studied the chemical potential dependence of
such states and compared it with the linear Kr\"onig-Penney 
excitation spectrum.

{\em Acknowledgements}
We thanks Lev Pitaevskii for several useful discussions.

\end{multicols}


\begin{references}

\bibitem{bronski01} J.C. Bronski, L.D. Carr, B. Deconinck, J.N. Kutz 
and K. Promislow, Phys. Rev. E {\bf 63} 036612 (2001).

\bibitem{niu1}  Biao Wu, Roberto B. Diener and Qian Niu, 
Phys. Rev. A {\bf 65} 25601 (2002).

\bibitem{niu2}  Biao Wu and Qian Niu, Phys. Rev. A {\bf 64} 61603(R) (2001).

\bibitem{muller}  E. J. Mueller, Phys. Rev. A {\bf 66}, 063603, (2002).

\bibitem{moler}  K. Berg-Soresen and K. Molmer, Phys. Rev. A {\bf 58}, 1480,
(1998).

\bibitem{Pethick}  M. Machholm, C. J. Pethick and H. Smith,
Phys. Rev. A {\bf 67}, 053613 (2003); M. Machholm, et al., Phys. Rev. A {\bf 69}, 043604 (2004).

\bibitem{pethick1}  D. Diakonov, L. M. Jensen, C. J. Pethick and H. Smith,
Phys. Rev. A {\bf 66}, 013604, (2002) 

\bibitem{stringari}  M. Kramer, L. Pitaevskii and S. Stringari, Phys. Rev.
Lett. {\bf 88} 180404, (2002).

\bibitem{stringari1}  M. Kramer, C. Menotti, L. Pitaevskii and S. Stringari, 
Eur. Phys. J. D {\bf 27}, 247 (2003).

\bibitem{optical} See, for instance, 
S. Flach and C.R. Willis, Phys. Rep. {\bf 295}, 182 (1998); P. G. Kevrekidis, K. $\Phi$. Rasmussen, A. R. Bishop, Int. J. Mod. Phys. B, {\bf 15} 2833 (2001) and ref.s therein.

\bibitem{hennig99} D.Hennig and G.P. Tsironis, Phys. Rep. 307, 333 (1999).

\bibitem{kittler}  C. Kittel, Introduction to solid state physics, (5th edition, 1976).

\bibitem{Reinhardt}  K. W. Mahmud, J. N. Kutz and W. P. Reinhardt,
cond-mat/0206532 (2002).

\bibitem{Reinhardt1}  L. D. Carr, C. W. Clark and W. P. Reinhardt, Phys.
Rev. A. {\bf 62}, 63610 (2000), ibid, {\bf 62}, 63611(2000).

\bibitem{Reinhardt2}  L. D. Carr, K. W. Mahmud and W. P. Reinhardt, Phys.
Rev. A. {\bf 64}, 33603 (2000).

\bibitem{presilla}  R. Dagosta and C. Presilla, Phys. Rev. A {\bf 65}, 043609
(2002).

\bibitem{raghavan99}  S. Raghavan, A. Smerzi, S. Fantoni, and S. R. Shenoy,
Phys. Rev. A {\bf 59}, 620 (1999).

\bibitem{gurevich} A. V. Gurevich and A. L. Krylov, 
Sov. Phys. JETP {\bf 65}, 5, (1987).

\bibitem{hand}  P. F. Byrd, M. D. Friedman, Handbook of Elliptic Integrals
for Engineers and Scientists, Second. Edition, Springer-Verlage New York
Heidelberg Berlin 1971.

\end{references}
\end{document}